\documentclass[reprint,onecolumn,superscriptaddress,citeautoscript,floatfix,longbibliography,bibnotes]{revtex4-2}
\usepackage{times,siunitx,amsmath,amsfonts,amsthm,amssymb,amsbsy,braket,graphicx,orcidlink}
\usepackage[utf8]{inputenc}
\usepackage[T1]{fontenc}
\usepackage[capitalize]{cleveref}
\newcommand{\ii}{\mathrm{i}}
\newcommand{\ee}{\mathrm{e}}

\renewcommand{\Im}{\mathrm{Im}}
\DeclareMathOperator{\sgn}{sgn}
\hypersetup{colorlinks=true, urlcolor=blue, linkcolor=blue, citecolor=red}
\begin{document}
%\title{1D Majorana Goldstinos and extended supersymmetry in quantum wires}
%\title{1D Majorana Goldstinos and partially-broken centrally-extended supersymmetry in quantum wires}
\title{1D Majorana Goldstinos and partial supersymmetry breaking in quantum wires}
\author{Pasquale Marra \orcidlink{0000-0002-9545-3314}}
\email{pmarra@ms.u-tokyo.ac.jp}
\affiliation{Graduate School of Mathematical Sciences, The University of Tokyo, 3-8-1 Komaba, Meguro, Tokyo, 153-8914, Japan}
\affiliation{Department of Physics, and Research and Education Center for Natural Sciences, Keio University, 4-1-1 Hiyoshi, Yokohama, Kanagawa, 223-8521, Japan}
\author{Daisuke Inotani \orcidlink{0000-0002-7300-1587}}
\author{Muneto Nitta \orcidlink{0000-0002-3851-9305}}
\affiliation{Department of Physics, and Research and Education Center for Natural Sciences, Keio University, 4-1-1 Hiyoshi, Yokohama, Kanagawa, 223-8521, Japan}
\date{\today}

\begin{abstract}
\paragraph*{Abstract}
%149 words
Realizing Majorana modes in topological superconductors, i.e., the condensed-matter counterpart of Majorana fermions in particle physics, may lead to a major advance in the field of topologically-protected quantum computation. Here, we introduce one-dimensional, counterpropagating, and dispersive Majorana modes as bulk excitations of a periodic chain of partially-overlapping, zero-dimensional Majorana modes in proximitized nanowires via periodically-modulated fields. This system realizes centrally-extended quantum-mechanical supersymmetry with spontaneous partial supersymmetry breaking. The massless Majorana modes are the Nambu-Goldstone fermions (Goldstinos) associated with the spontaneously broken supersymmetry. Their experimental fingerprint is a dip-to-peak transition in the zero-bias conductance, which is generally not expected for Majorana modes overlapping at a finite distance. Moreover, the Majorana modes can slide along the wire by applying a rotating magnetic field, realizing a ``Majorana pump''. This may suggest new braiding protocols and implementations of topological qubits.
%One-dimensional Majorana modes can be obtained as boundary excitations of topologically nontrivial two-dimensional topological superconductors. Here, we propose instead the bottom-up creation of one-dimensional, counterpropagating, and dispersive Majorana modes as bulk excitations of a periodic chain of partially-overlapping, zero-dimensional Majorana modes in proximitized quantum nanowires via periodically-modulated magnetic fields with large amplitude variations. These Majorana modes are pseudohelical, having opposite Majorana pseudospin, and realize, for the first time in condensed matter systems,  a centrally-extended quantum-mechanical supersymmetry with spontaneous partial supersymmetry breaking. We identify the massless Majorana fermions as Goldstinos, i.e., the Nambu-Goldstone fermions associated with the spontaneous breaking of supersymmetry. The experimental fingerprint of supersymmetry is the presence of a zero-bias peak, which is generally not expected for overlapping Majorana modes localized at a finite distance. Moreover, slowly varying magnetic fields induce an adiabatic Majorana pump which can be used as a dynamical probe of topological superconductivity.
\end{abstract}
\maketitle

\section*{Introduction}

Majorana fermions in high-energy physics are spin-$1/2$ particles that are symmetric with respect to charge conjugation symmetry, i.e., neutral fermions that coincide with their own anti-particles~\cite{majorana_teoria_1937,wilczek_majorana_2009,elliott_colloquium_2015}.
In condensed matter, they appear as quasiparticle excitations in superconductors, where particle-hole symmetry plays the role of charge conjugation~\cite{alicea_new_2012,leijnse_introduction_2012, beenakker_search_2013,elliott_colloquium_2015,sato_majorana_2016,sato_topological_2017,aguado_majorana_2017,haim_timereversalinvariant_2019}.
Generally, Majorana quasiparticles are topologically protected $(d-1)$-dimensional boundary excitations of a topologically nontrivial $d$-dimensional bulk.
Specifically, 0-dimensional (0D) Majorana modes~\cite{kitaev_unpaired_2001,oreg_helical_2010,lutchyn_majorana_2010} correspond to the end states of 1D quantum systems with proximitized superconductivity, whereas chiral and helical 1D Majorana modes correspond to the edge states of 2D unconventional superconductors or planar superconducting heterostructures~\cite{qi_timereversalinvariant_2009,qi_chiral_2010,nakosai_topological_2012,seradjeh_majorana_2012,zhang_timereversalinvariant_2013,sun_helical_2016,chen_helical_2018,he_platform_2019,hu_chiral_2019,hogl_chiral_2020,shen_spectroscopic_2020}, respectively with broken or unbroken time-reversal symmetry.

Majorana quasiparticles exhibit remarkable properties such as non-abelian statistics~\cite{ivanov_nonabelian_2001,alicea_nonabelian_2011}, conformal invariance~\cite{sato_majorana_2016}, and emergent supersymmetry (SUSY)~\cite{witten_dynamical_1981,cooper_supersymmetry_1995,gangopadhyaya_supersymmetric_2017,grover_emergent_2014,qi_time-reversal-invariant_2009,hsieh_all_2016,huang_supersymmetry_2017,rahmani_emergent_2015,rahmani_phase_2015,rahmani_interacting_2019}.
Besides their purely theoretical appeal, Majorana quasiparticles attracted enormous interest due to their potential technological applications in quantum computing~\cite{ivanov_nonabelian_2001,kitaev_faulttolerant_2003,nayak_nonabelian_2008,alicea_nonabelian_2011,sarma_majorana_2015,aasen_milestones_2016,karzig_scalable_2017,lahtinen_short_2017,lian_topological_2018}.
Quite a few experiments observed signatures compatible with the presence of spatially-separated 0D Majorana modes in nanowires~\cite{mourik_signatures_2012,lee_zerobias_2012,rokhinson_fractional_2012,das_zerobias_2012,deng_anomalous_2012,finck_anomalous_2013,churchill_superconductor_2013,lee_spinresolved_2014,deng_majorana_2016,nichele_scaling_2017,chen_experimental_2017,gul_ballistic_2018,grivnin_concomitant_2019} and quantum chains of adatoms~\cite{choy_majorana_2011,pientka_topological_2013,nadj-perge_observation_2014,pawlak_probing_2016,feldman_highresolution_2017,kim_tailoring_2018,pawlak_majorana_2019}, and chiral 1D Majorana modes in planar heterostructures~\cite{he_chiral_2017,menard_two-dimensional_2017,palacio-morales_atomicscale_2019,kayyalha_absence_2020,shen_spectroscopic_2020}.
Similar experimental signatures can be reproduced, however, by trivial Andreev bound states or by Majorana modes localized at a finite distance and with a finite overlap, known as quasi-Majorana modes~\cite{prada_andreev_2020}, in the presence of inhomogeneous potentials~\cite{kells_nearzeroenergy_2012,stanescu_disentangling_2013,liu_andreev_2017,liu_distinguishing_2018,moore_quantized_2018,marra_majorana/andreev_2022}.

\begin{figure}[t]
\centering
\includegraphics[width=\columnwidth]{Fig1.pdf}
\caption{
Lattice of 0D Majorana modes in a proximitized nanowire.
\\
A semiconducting nanowire with Rashba spin-orbit coupling $\alpha$ covered by a superconducting shell, with a periodically-modulated magnetic field induced by a regular array of nanomagnets and an externally applied field $B_\text{a}$.
The periodic modulation of the field induces topologically nontrivial (NT) segments alternating to trivial ones, which correspond respectively to negative and positive values of the local Majorana mass $\mathcal M$.
0-dimensional quasi-Majorana modes $\gamma_{Aj}$ and $\gamma_{Bj}$ localize at the boundaries between trivial and nontrivial segments, with overlaps $w$ and $v$.
}
\label{fig:system}
\end{figure}

%In this work, we propose a \emph{bottom-up} approach to topological superconductivity, i.e., employing a periodic array of partially-overlapping 0D quasi-Majorana modes to realize 1D Majorana fermions with emergent SUSY\@.
In this work, we propose the realization of centrally extended quantum-mechanical SUSY, i.e., extended SUSY with central charges, in an experimentally-accessible condensed matter system, by employing a periodic array of partially-overlapping 0D quasi-Majorana modes to realize a Majorana chain with dispersive 1D Majorana fermions, and show how this can be achieved in proximitized semiconducting nanowires~\cite{stanescu_majorana_2013,lutchyn_majorana_2018,zhang_next_2019,frolov_topological_2020,flensberg_engineered_2021} via periodically-modulated magnetic fields~\cite{klinovaja_transition_2012,kjaergaard_majorana_2012,ojanen_majorana_2013,maurer_designing_2018,marra_controlling_2017,marra_topologically_2019} with large variations of the field intensity.
We find that, for strong variations of the magnetic field amplitude, the topological mass gap $\mathcal M$ can assume alternatively positive and negative values along the wire, corresponding to topologically trivial and nontrivial segments, with partially-overlapping 0D quasi-Majorana modes localized at their boundaries and forming a 1D periodic lattice.
This is the first realistic proposal to realize the Majorana chain model~\cite{rahmani_phase_2015,rahmani_emergent_2015,hsieh_all_2016,huang_supersymmetry_2017,sannomiya_supersymmetry_2019} in proximitized nanowires with strong spin-orbit coupling.
The system exhibits a pair of dispersive and counterpropagating 1D Majorana modes delocalized along the whole wire and separated from the higher-energy bulk states, with a mass gap that can be tuned by externally applied fields.
To characterize these emergent 1D Majorana modes, we introduce the concept of pseudohelicity and unveil the existence of an extended SUSY algebra with central charges.
In the massless case, we found indeed that the 1D Majorana modes are pseudohelical, i.e., have opposite Majorana pseudospin, and exhibit centrally extended SUSY, with a finite zero-energy density of states and zero-bias peak delocalized along the whole wire.
We find that the signature of the emergent SUSY can be revealed by the transition from a dip $G=0$ to a quantized peak $G=2e^2/h$ in the zero-bias conductance, or by inducing an adiabatic Majorana pumping in a sliding lattice of 0D quasi-Majorana modes, with quantized transport of one quasi-Majorana mode per a half cycle.
Note that zero-bias peaks are generally not expected in the presence of several 0D quasi-Majorana modes localized at a finite distance.
Moreover, we identify the massless Majorana fermion with a Goldstino, i.e., the Nambu-Goldstone fermion~\cite{volkov_is_1973} associated with spontaneously broken SUSY from ${\mathcal N}=4$ to ${\mathcal N}=2$.
Such a partial SUSY breaking of the extended SUSY is known to be possible only in the presence of central charges.
This is perfectly compatible with our setup, in which we explicitly identify the central charges of the extended superalgebra.
While the extended SUSY was proposed in several condensed matter systems~\cite{santachiara_supersymmetric_2005,hagendorf_open_2017,behrends_supersymmetry_2020}, to our knowledge, this is the first real-world realization of centrally-extended quantum-mechanical SUSY, which plays essential roles in non-perturbative aspects of quantum field theory in high energy physics~\cite{Witten:1978mh,Seiberg:1994rs,Seiberg:1994aj}, and partial breaking of the extended SUSY algebras.

\section*{Results}

\subsection*{Effective model: 1D lattice of 0D Majorana modes}

We consider a bipartite 1D lattice of $2N$ 0D Majorana modes
\begin{equation}\label{eq:H}
\mathcal{H}_\text{eff}=\ii \sum_{j=1}^N
\left(
w \gamma_{Aj} \gamma_{Bj} + v \gamma_{Bj} \gamma_{Aj+1}
\right),
\end{equation}
where $\gamma_{Aj},\gamma_{Bj}$ are the Majorana operators corresponding to a single Dirac operator $c_j=(\gamma_{Bj}+\ii\gamma_{Aj})/2$ per unit cell, and with $w,v\in\mathbb R$.
This model is a special case of the Kitaev chain model~\cite{kitaev_unpaired_2001} if $\mu =2w$ and $t = \Delta =-v$.
In momentum space we get
\begin{equation}
\mathcal{H}_\text{eff}=
\sum_k
[c_k^\dagger , c_{-k}]
\cdot
\mathbf{H}_\text{eff}(k) \cdot \boldsymbol \tau
\cdot
\begin{bmatrix}
c_k \\
c_{-k}^\dagger\\
\end{bmatrix}
,
\end{equation}
up to a constant term, with $\mathbf{H}_\text{eff}(k) =\left(0, v\sin{k}, v\cos{k} -w\right)$, and $\boldsymbol\tau$ the vector of Pauli matrices.
The energy dispersion is
\begin{equation}
E_k=|\mathbf{H}_\text{eff}(k)|=\sqrt{w^2 + v^2 - 2 w v \cos{k}},
\end{equation}
with a topological mass gap $\mathcal M_\text{eff}= |w|-|v|$.
In the continuum limit (and assuming $vw>0$), the Hamiltonian coincides with a 1D Dirac equation $H= v k \,\tau_y + \left(m v^2-\frac{v}2 k^2\right)\tau_z$ with mass gap $m v^2=v-w$ and a quadratic correction in the momentum.
The covariant form is obtained by multiplying the Hamiltonian by $\tau_z$.
The Dirac equation is topologically trivial or nontrivial, respectively, for $mv<0$ and $mv>0$, i.e., for $\mathcal M_\text{eff}>0$ and $\mathcal M_\text{eff}<0$~\cite{shen_topological_2011}.

In the massless case $|v|=|w|$ the zero-energy eigenstates are doubly degenerate at gapless points and described by the fermionic operator $d_\text{M}=({\widetilde\gamma}_A+\ii{\widetilde\gamma}_B)/2$ and its hermitian conjugate $d_\text{M}^\dag$, where the nonlocal Majorana operators are ${\widetilde\gamma}_A=(1/{\sqrt{N}})\sum_j\gamma_{Aj}$, and ${\widetilde\gamma}_B=(1/{\sqrt{N}})\sum_j\gamma_{Bj}$.
The gapless state $v=\pm w$ separates two topologically inequivalent phases described by the topological invariant $\sgn{\mathcal M_\text{eff}}=\pm 1$ where $\mathcal M_\text{eff}=|w|-|v|$.
The Hamiltonian also exhibits 0D Majorana end modes in the nontrivial phase $|v|>|w|$ in the case of open boundary conditions.
The two end states are ${\widetilde\gamma}_\text{L}\propto\sum_j(w/v)^j \gamma_{Aj}$ and ${\widetilde\gamma}_\text{R}\propto\sum_j(w/v)^{N+1-j} \gamma_{Bj}$ localized at the opposite ends of the chain with localization length $\xi_\text{eff}=1/|\log|w/v||$.

The massless Majorana fields ($|v|=|w|$) describe a 1D free Majorana fermion in a 1+1D conformal field theory~\cite{sato_majorana_2016}, which coincides with a pair of counterpropagating 1D Majorana modes.
The solutions of the Dirac equation for a relativistic massless particle $m=0$ form a helical pair of counterpropagating modes with opposite spin.
However, our effective model is spinless:
The role played by the spin degree of freedom is now played by the particle-hole degree of freedom.
Hence, to characterize the properties of the 1D Majorana modes in our model, we introduce the Majorana pseudospin operator as $\boldsymbol\tau/2$ in analogy to the spin operator $\boldsymbol\sigma/2$ (we use natural units).
We obtain that the expectation values of the Majorana pseudospin
\begin{equation}
\langle\boldsymbol\tau/2\rangle=\frac{\mathbf{H}_\text{eff}(k)}{2E_k},
\end{equation}
have opposite directions for the two modes near the gapless point, being $\langle\boldsymbol\tau\rangle =\sgn{(v\sin{k})}\hat{\mathbf{y}}$ at $k\to0,\pi$ for $v=\pm w$, respectively.
Analogously to the notion of helical modes~\cite{qi_timereversalinvariant_2009}, i.e., a pair of counterpropagating modes having opposite spin, we introduce the notion of pseudohelical modes, as a pair of counterpropagating modes having opposite Majorana pseudospin.
Hence, the two modes form a pseudohelical pair near the gapless point.
In this case, elastic backscattering is suppressed since the two modes crossing at zero energy are orthogonal, analogously to the case of helical modes~\cite{qi_timereversalinvariant_2009}.
The Majorana pseudospin introduced here generalizes the Majorana polarization~\cite{sticlet_spin_2012,marra_majorana/andreev_2022}:
The expectation values of the two components $\langle\boldsymbol\tau_{x,y}/2\rangle$ coincide with the Majorana polarization (up to prefactors).

\subsection*{Centrally extended superalgebra and partial supersymmetry breaking}

In the massless case, we find that \cref{eq:H} exhibits extended $\mathcal N=4$ quantum mechanical SUSY given by the combined algebra defined by the supercharges
\begin{subequations}\label{eq:supercharges}
\begin{align}\label{eq:supercharge1}
{\mathcal Q}_1&=\sqrt{\frac{{\mathcal H}_\text{SUSY}}2}\, d_\text{M} (1 + P),
\\\label{eq:supercharge2}
{\mathcal Q}_2&=\sqrt{\frac{{\mathcal H}_\text{SUSY}}2}\, T (1 + P),
\end{align}
\end{subequations}
which satisfy the superalgebra
$\{{\mathcal Q}_i, {\mathcal Q}_j^\dag\}=2\delta_{ij}{\mathcal H}_\text{SUSY}+{\mathcal Z}_{ij}$,
$\{{\mathcal Q}_i,{\mathcal Q}_j\}=\{{\mathcal Q}_i^\dag,{\mathcal Q}_j^\dag\}=0$,
$\{P, {\mathcal Q}_i \}=0$,
with central charges
\begin{subequations}\label{eq:centralcharges}
\begin{align}
{\mathcal Z}_{11}&=
-{\mathcal H}_\text{SUSY}
(1+P (-1)^{d^\dag_\text{M}d_\text{M}} ),
\\
{\mathcal Z}_{22}&=0,
\\
{\mathcal Z}_{12}={\mathcal Z}_{21}^\dag&= {\mathcal H}_\text{SUSY}\{ d_\text{M} (1+P), T^\dag \}.
\end{align}
\end{subequations}
Here, ${\mathcal H}_\text{SUSY}={\mathcal H}_\text{eff} + 2h|v|$ (with $h>1$) is the many-body Hamiltonian having nonnegative energy levels, $P=\prod_{j=1}^N \ii \gamma_{Aj}\gamma_{Bj}$ the fermion parity, and $T$ the translation defined by $T \gamma_{Aj} T^\dag= \gamma_{Bj}$, $T \gamma_{Bj} T^\dag= \gamma_{Aj+1(\mathrm{mod} N)}$, which satisfies $\{T,P\}=0$ and $[T,{\mathcal H}_\text{SUSY}]=0$ for $|v|=|w|$.
(The supercharges ${\mathcal Q}_2$ and ${\mathcal Q}_1$ were introduced separately in previous works~\cite{hsieh_all_2016,huang_supersymmetry_2017}, but the existence of the combined $\mathcal N=4$ superalgebra has not been previously demonstrated in Majorana chain models, up to our knowledge.)
All many-body eigenstates, including the groundstate, have superpartners with opposite parity.
Thus, the Witten index is zero, and SUSY is spontaneously broken~\cite{witten_dynamical_1981}.
The two degenerate groundstates are the vacuum $\ket{0}$ and the state $\ket{1}=d_\text{M}^\dag\ket{0}$, respectively with even and odd fermion parity.
The supercharge ${\mathcal Q}_1\ket{0}$ annihilates both groundstates
${\mathcal Q}_1\ket{0} = {\mathcal Q}_1\ket{1} = {\mathcal Q}_1^\dag\ket{0}={\mathcal Q}_1^\dag\ket{1}=0$.
However, the two groundstates are superpartners with respect to ${\mathcal Q}_2$:
Since $[T,{\mathcal H}_\text{SUSY}]=0$ and $\{T,P\}=0$, the eigenstates $\ket{0}$ and $T\ket{0}$ have same energy but opposite parities, which mandates $T\ket{0}=\ket{1}$ and $T\ket{1}=\ket{0}$ (up to a complex phase).
This yields
\begin{equation}\label{eq:partiallybroken}
{\mathcal Q}_2\ket{0} =
\sqrt{2{\mathcal H}_\text{SUSY}}\, \ket{1},
\qquad
{\mathcal Q}_2^\dag\ket{1} =
\sqrt{2{\mathcal H}_\text{SUSY}}\, \ket{0},
\qquad
{\mathcal Q}_2^\dag\ket{0}=
{\mathcal Q}_2\ket{1}=0.
\end{equation}
Hence, the supersymmetry ${\mathcal Q}_1$ is unbroken whereas ${\mathcal Q}_2$ is spontaneously broken:
The $\mathcal N=4$ superalgebra $({\mathcal Q}_1,{\mathcal Q}_2)$ is spontaneously broken down into the $\mathcal N=2$ superalgebra ${\mathcal Q}_1$.
This mandates the presence of a Goldstino~\cite{volkov_is_1973}, which we identify with the massless Majorana fermion.
The zero mass gap is protected by SUSY, i.e., the gap closes if and only if the Hamiltonian exhibits SUSY\@.
We note that the supercharge ${\mathcal Q}_1$ can be defined even in the presence of disorder and broken translational symmetry~\cite{huang_supersymmetry_2017,longpaper}:
In this case, the corresponding Goldstino has a mass gap which closes if and only if the Hamiltonian exhibits SUSY\@.
We note that the no-go theorem by Witten~\cite{witten_dynamical_1981,witten_constraints_1982} forbids partial supersymmetry breaking in extended superalgebras with zero central charges:
Either all supersymmetries $\mathcal{Q}_i$ are broken, or they are all unbroken.
However, the no-go theorem can be evaded in the presence of nonzero central charges~\cite{ivanov_partial_1991}, which allow the partial breaking of the extended SUSY.

\subsection*{Proximitized semiconducting nanowire with spin-orbit coupling and periodic magnetic field}

The Hamiltonian in \cref{eq:H} may describe the low-energy effective theory of a 1D topological superconductor with spatially-modulated fields.
Specifically, we consider a semiconducting nanowire with Rashba spin-orbit coupling and coated with a conventional superconductor, as in \cref{fig:system}.
A periodically-modulated magnetic field in the $zx$ plane $\mathbf{B}_\text{nm}(x)$ is induced by an array of nanomagnets~\cite{klinovaja_transition_2012,kjaergaard_majorana_2012,ojanen_majorana_2013,maurer_designing_2018,kornich_majorana_2020} with magnetic moments parallel to the $z$-axis.
In addition, we consider a uniform applied field $B_\text{a}$ in the $z$-direction, which can be used to control the lengths of the nontrivial and trivial segments.
The wire is described by the Hamiltonian density
\begin{equation}\label{eq:LEH}
H= \left( \frac{p^2}{2m} + \frac{\alpha}{\hbar} \sigma_y p - \mu \right) \tau_z
+ \mathbf{b}(x) \cdot \boldsymbol{\sigma}
+ \Delta(x) \tau_x,
\end{equation}
where $\boldsymbol\sigma$ and $\boldsymbol\tau$ are the vectors of Pauli matrices in spin and particle-hole space, $m$ the effective mass, $\alpha$ the spin-orbit coupling, $\mu$ the chemical potential, $\mathbf b(x)=(g\mu_\text{B}/2) \mathbf{B}(x)$ the Zeeman field in the $zx$ plane ($\mathbf{B}(x)=\mathbf{B}_\text{nm}(x)+B_\text{a} \hat{\mathbf{z}}$ is the total magnetic field), and $\Delta(x)$ the proximization-induced superconducting pairing.
We require the wavelength $\lambda$ of the periodically-modulated field to be comparable to the Majorana localization length, which is $\xi_\text{M}\approx (b/E_\text{SO})\alpha/\Delta$ and $\alpha/\Delta$ respectively for $E_\text{SO}=m\alpha^2/2\hbar^2\ll\Delta$ and $\gg\Delta$ (weak and strong spin-orbit coupling regimes)~\cite{klinovaja_composite_2012,mishmash_approaching_2016,aguado_majorana_2017}.
It is essential to our proposal to consider magnetic fields with large variations of the field intensity, contrarily to the well-known regime of periodically rotating fields with constant amplitude (or negligible amplitude variations), considered before~\cite{braunecker_spinselective_2010,kjaergaard_majorana_2012,klinovaja_transition_2012,klinovaja_topological_2013}.
Notice that periodically-modulated magnetic fields can also be induced by employing magnetic textures~\cite{mohanta_electrical_2019,desjardins_synthetic_2019} or domain walls~\cite{neupert_chain_2010,rex_majorana_2020}, while periodically-modulated chemical potentials can be obtained by a periodic modulation of the width of the superconducting coating~\cite{woods_enhanced_2020,longpaper} in epitaxial 1D semiconductor-superconductor heterostructures~\cite{shabani_two-dimensional_2016,hell_two-dimensional_2017,pientka_topological_2017,suominen_zero-energy_2017}.

\begin{figure}[t]
\centering
\includegraphics[width=.95\columnwidth]{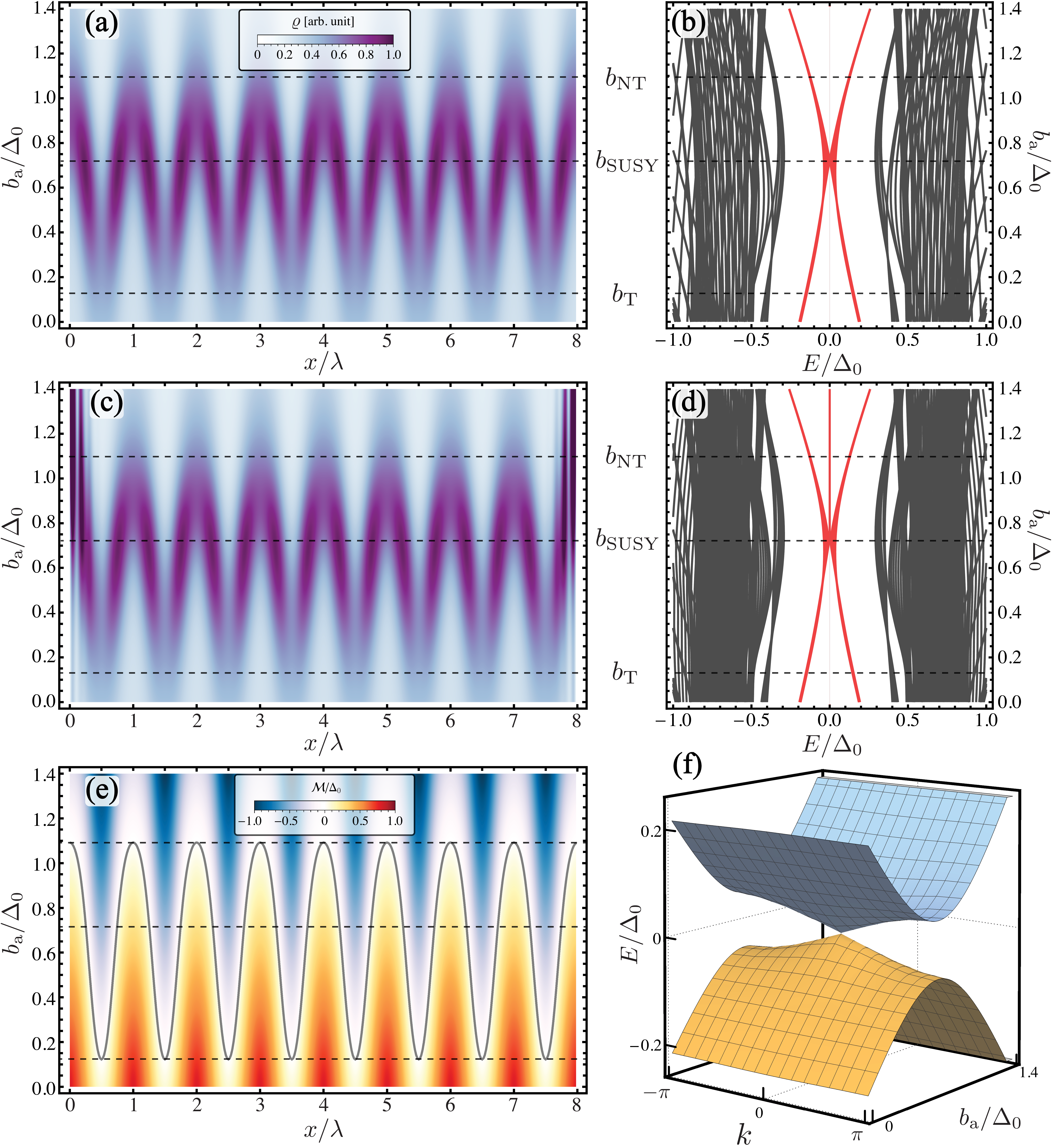}
\caption{
Local density of states and energy spectra of a Majorana lattice.
\\
Numerically calculated local density of states (LDOS) and energy spectra as a function of the applied magnetic field intensity, showing a 1D lattice of 0D quasi-Majorana modes.
(a)
LDOS at zero energy of a nanowire in a periodically-modulated magnetic field, as a function of the position $x$ and the external field $b_\text{a}$ applied in the $z$-direction, calculated with periodic boundary conditions.
The peaks of the LDOS indicate the presence of 1D Majorana modes, corresponding to a periodic lattice of overlapping 0D quasi-Majorana modes localized at the boundaries between trivial ($\mathcal M>0$) and nontrivial ($\mathcal M<0$) segments.
(b)
Energy spectra $E$ with dispersive 1D Majorana modes (highlighted) below the particle-hole gap.
The dispersion becomes gapless (massless) when the overlaps between localized 0D quasi-Majorana modes across the nontrivial and trivial segments become equal when $b_\text{a}= b_\text{SUSY}$.
(c) and (d)
Same as before, but with open boundary conditions.
The wire becomes nontrivial after the closing of the particle-hole gap for $b_\text{a}>b_\text{SUSY}$ and exhibits 0D quasi-Majorana modes localized at its opposite ends with diverging density (out of scale).
(e)
Local Majorana mass and its nodes $\mathcal M=0$ (continuous line) as a function of the position and applied field.
When $b_\text{a}>b_\text{NT}$, the Majorana mass becomes negative on the whole wire and the trivial segments disappear.
Conversely, when $b_\text{a}<b_\text{T}$, the Majorana mass becomes positive on the whole wire and the nontrivial segments disappear.
(f)
Energy dispersion of the 1D Majorana modes as a function of the momentum $k$ and the applied field.
%The LDOS integrated over the whole energy dispersion of the 1D Majorana mode is shown in Supplementary Figure 3.
%The LDOS as a function of the energy is shown in the Supplementary Movie.
Energies and magnetic fields are in units of the superconducting gap $\Delta_0$, lengths are in units of the wavelength $\lambda$ of the periodically-modulated field.
}
\label{fig:energy}
\end{figure}

If all fields are uniform, the Hamiltonian above reduces to the Oreg-Lutchyn minimal model~\cite{oreg_helical_2010,lutchyn_majorana_2010}.
The sign of the Majorana mass gap $\mathcal M=\sqrt{\mu^2+\Delta^2}-|b|$ characterize the trivial ($\mathcal M>0$) and nontrivial phases ($\mathcal M<0$) with $\mathcal M=0$ at the closing of the particle-hole gap.
If the chemical potential, Zeeman field, or superconducting pairing is not uniform, one can define a local Majorana mass gap $\mathcal M(x)=\sqrt{\mu(x)^2+\Delta(x)^2}-|b(x)|$ which may be alternatively positive and negative values along the wire.
In this case, segments with $\mathcal M>0$ and $\mathcal M<0$ are trivial and nontrivial with a local topological invariant ${\mathcal P(x)}=\sgn{\mathcal M(x)}$.
Hence, 0D quasi-Majorana modes localize at the boundaries between trivial and nontrivial segments at the nodes of the Majorana mass gap $\mathcal M(x)=0$ (see \cref{fig:system}) with localization length $\xi_\text{M}$ and mutual distance $L_{AB}$ and $L_{BA}$.
If the wavelength $\lambda$ of the periodically-modulated field is comparable with the Majorana localization length $\xi_\text{M}$, the 0D quasi-Majorana mode $\gamma_{Aj}$ and $\gamma_{Bj}$ have finite overlaps $w\propto \ee^{-L_{AB}/\xi_\text{M}}$ and $v\propto \ee^{-L_{BA}/\xi_\text{M}}$, and realize a periodic bi-partite 1D lattice.
Notice that the values of the overlaps $v,w$ depend strongly on the distance between the nodes of the Majorana mass, i.e., the distance between neighboring 0D quasi-Majorana modes.
Hence, projecting onto the subspace of Majorana operators, one obtains the effective low-energy Hamiltonian in \cref{eq:H}, where the coupling parameters $v,w$ coincide with the Hamiltonian matrix elements between contiguous 0D quasi-Majorana modes separated by trivial and nontrivial segments, respectively.
For $|v|=|w|$, the overlaps between contiguous 0D quasi-Majorana modes across trivial and nontrivial segments become equal, and the 1D Majorana modes become massless (gapless).
We calculate the magnetic field $\mathbf{B}_\text{nm}$ of the nanomagnets array via the finite-element method, see Supplementary Figure 1.
We use the resulting Zeeman field in all numerical calculations.
For reference, we find that a reasonable approximation of the Zeeman field is
\begin{equation}\label{eq:field}
 \mathbf{b}_\text{nm}(x) \approx
 {b}_\text{nm}
 \left[
 \left(1-\cos{(2\pi x/\lambda)}\right)\hat{\mathbf{z}}
 - \sin{(2\pi x/\lambda)}\hat{\mathbf{x}}
 \right]
 ,
\end{equation}
where ${b}_\text{nm}$ coincides with the magnitude of the average field along the wire, which gives
$|\mathbf{b}_\text{nm}(x)| \approx |{b}_\text{nm}| \sqrt{\left(2-2\cos{(2\pi x/\lambda)}\right)}$.
%$$ \mathcal{M}(x)\approx\sqrt{\mu^2+\Delta^2}-|{b}_\text{nm}| \sqrt{\left(2-2\cos{(2\pi x/\lambda)}\right)}$$
%$$|\mathbf{b}_\text{nm}(x)|^2 \approx 2|{b}_\text{nm}|^2 \left(1-\cos{(2\pi x/\lambda)}\right)$$
We then discretize the Hamiltonian and take realistic values for the model parameters~\cite{mourik_signatures_2012,liu_andreev_2017}.
Moreover, we assume $\Delta(x)\propto \Delta_0\sqrt{1-B(x)^2/B_\text{c}^2}$, where $\Delta_0$ is the superconducting gap at zero field and $B_\text{c}$ the critical field, in order to account for the suppression of the superconducting pairing induced by the magnetic field.
By numerically diagonalizing the discretized Hamiltonian, we obtain the energy $E_n$ and Nambu spinor $\Psi_n(x)$ of each eigenstate (see Methods section).

Figures~\ref{fig:energy}(a) and \ref{fig:energy}(b) show the local density of states (LDOS) at zero energy
$\varrho(x)=\frac1\pi \Im \sum_{n} {|\Psi_{n}(x)|^2}/{(E_n-\ii\Gamma)}$
with finite broadening $\Gamma=0.1\Delta_0$ (to simulate the experimental conditions), and energy spectra as a function of the uniform applied magnetic field, in the case of periodic boundary conditions.
The LDOS shows the presence of a periodic lattice of 0D quasi-Majorana modes with finite overlap, localized at the nodes of the Majorana mass gap $\mathcal M(x)=0$ [see also \cref{fig:energy}(e)].
This lattice corresponds to dispersive 1D Majorana modes with energy below the particle-hole gap and separated from the higher-energy bulk states, highlighted in \cref{fig:energy}(b), and with a Majorana mass gap equal to $\mathcal M_\text{eff}=|w|-|v|$.
When the overlaps $v,w$ between 0D quasi-Majorana modes across the nontrivial and trivial segments become equal when $b_\text{a}= b_\text{SUSY}$ (which, in first approximation, occurs when $L_{AB}\approx L_{BA}$), the periodic lattice becomes invariant up to translations $T$ [see \cref{eq:supercharges}].
Hence, the wire exhibits SUSY and the dispersion becomes gapless (massless) with maximum LDOS at zero energy.

When the applied field increases above the threshold $b_\text{NT}$, such that $|b(x)|\ge\sqrt{\mu(x)^2+\Delta(x)^2}$ $\forall x$, the Majorana mass becomes negative on the whole wire and the trivial segments disappear.
Conversely, when the applied field decreases below the threshold $b_\text{T}$, such that $|b(x)|\le\sqrt{\mu(x)^2+\Delta(x)^2}$ $\forall x$, the Majorana mass becomes positive on the whole wire and the nontrivial segments disappear.
In these two cases, the 0D quasi-Majorana modes at the ends of the trivial (or nontrivial) segments fuse into finite-energy Andreev-like fermionic modes.
The continuous crossover between Majorana and Andreev-like modes is realized by increasing the overlaps between contiguous 0D quasi-Majorana modes at the ends of either the trivial or nontrivial segments, such that $|v|\gg|w|$ or $|w|\gg|v|$, without closing the particle-hole gap.
This crossover also occurs when the wavelength $\lambda$ becomes smaller than the Majorana localization length $\lambda\lesssim\xi_\text{M}$:
This results in larger overlaps $v,w$ between contiguous 0D quasi-Majorana modes fusing into fermionic Andreev-like modes.
In the opposite regime $\lambda\gg\xi_\text{M}$ one has $v,w\to0$, which corresponds to decoupled quasi-Majorana modes with flat dispersion $E_k\approx0$.

Figures~\ref{fig:energy}(c) and \ref{fig:energy}(d) show the LDOS and the energy spectra in the case of open boundary conditions.
For $b_\text{a}<b_\text{T}$ and $b_\text{a}>b_\text{NT}$ the local Majorana mass gap $\mathcal M(x)$ has the same sign along the wire, realizing a gapped trivial or nontrivial phase.
In the latter case, 0D Majorana end modes ${\widetilde\gamma}_\text{L}$, ${\widetilde\gamma}_\text{R}$ localize at the opposite ends.
However, for $b_\text{T}<b_\text{a}<b_\text{NT}$, the local Majorana mass gap $\mathcal M(x)$ changes its sign along the wire, with 0D quasi-Majorana modes described by the effective Hamiltonian in \cref{eq:H}.
This effective Hamiltonian can be trivial or nontrivial with Majorana mass gap $\mathcal{M}_\text{eff}=|w|-|v|$, as analyzed before.
Hence, the nontrivial phase with 0D Majorana end modes ${\widetilde\gamma}_\text{L}$, ${\widetilde\gamma}_\text{R}$ is also realized for $b_\text{SUSY}<b_\text{a}<b_\text{NT}$ (i.e., $|v|>|w|$).
The LDOS integrated over the whole energy dispersion of the 1D Majorana mode is shown in Supplementary Figure 3.
The LDOS as a function of energy below the bulk gap is shown in the Supplementary Movie.

\Cref{fig:energy}(e) shows the local Majorana mass $\mathcal M(x)$ as a function of the position and the applied magnetic field.
For reference, we draw a continuous line at the boundary between trivial and nontrivial phases at $\mathcal M=0$, where the 0D quasi-Majorana modes localize.
\Cref{fig:energy}(f) shows the momentum dispersion of the 1D Majorana modes calculated numerically as a function of the applied field, which corresponds to the highlighted subgap state in \cref{fig:energy}(b).
The 1D Majorana modes correspond to a periodic lattice of localized and overlapping 0D quasi-Majorana modes.
The mass gap of the 1D Majorana modes closes when $b_\text{a}= b_\text{SUSY}$.

\subsection*{Sliding lattice and Majorana pump}

\begin{figure}[t]
\centering
\includegraphics[width=.95\columnwidth]{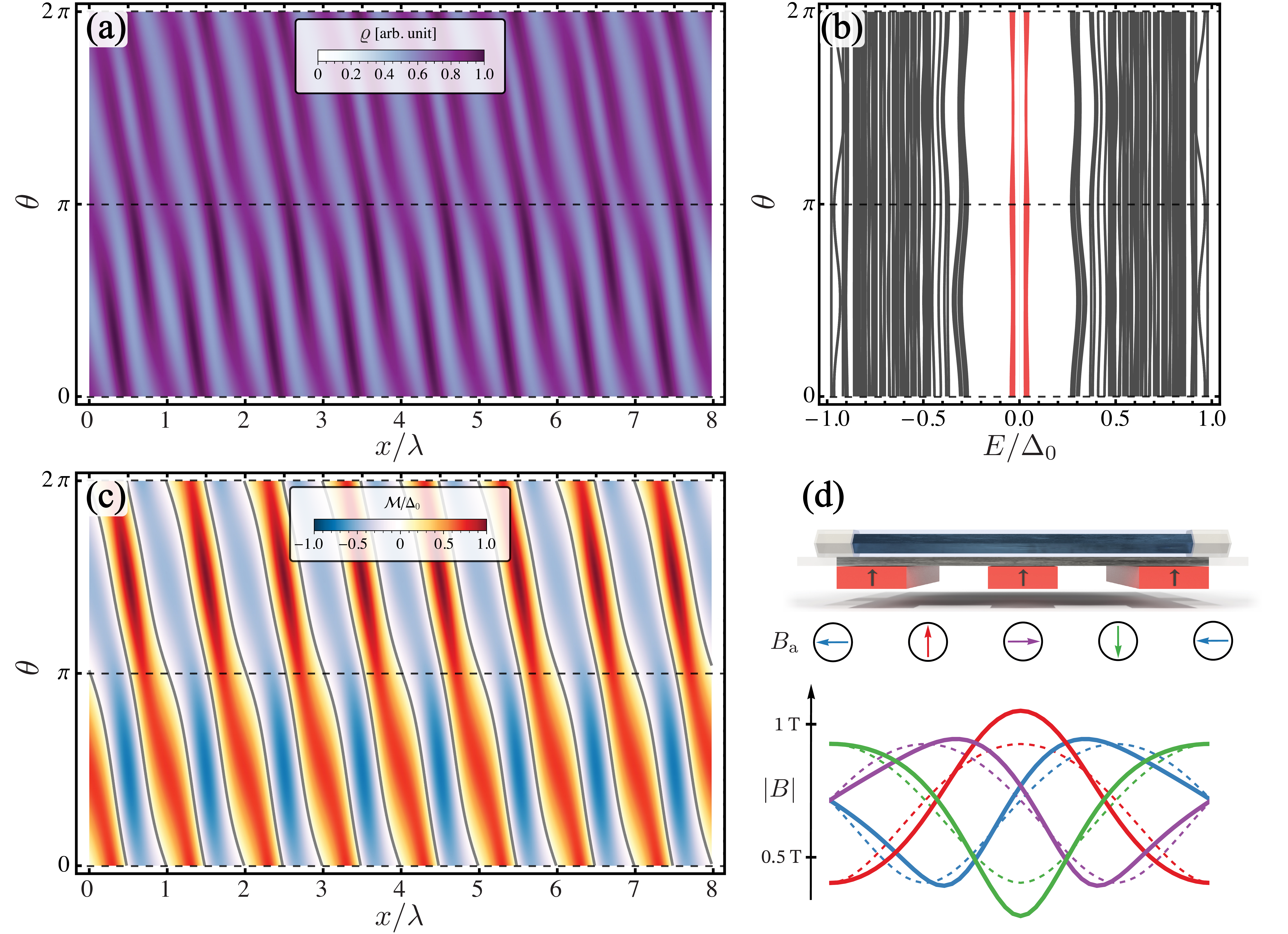}
\caption{
Local density of states and energy spectra of a sliding Majorana  lattice.
\\
Numerically calculated local density of states (LDOS) and energy spectra as a function of the applied magnetic field direction, showing a sliding 1D lattice of 0D quasi-Majorana modes.
(a)
LDOS at zero energy as a function of the position $x$ and the magnetic field direction $\theta$ in the $zx$-plane, calculated with closed boundary conditions.
0D quasi-Majorana modes localize at the boundaries between trivial ($\mathcal M>0$) and nontrivial ($\mathcal M<0$) segments.
As the field rotates, the lattice of 0D quasi-Majorana modes slides in the $x$-direction.
(b)
Energy spectra $E$ with dispersive 1D Majorana modes (highlighted) below the gap.
(c)
Local Majorana mass $\mathcal M$ as a function of the applied field direction $\theta$, with nodes at $\mathcal M=0$ (continuous lines).
(d)
Total magnetic field calculated as the superposition of the applied field $B_\text{a}$ and the nanomagnets fields for different directions $\theta$ of the applied field (continuous lines) compared with a sliding harmonic field $\propto\sin{(2\pi x/\lambda+\theta)}$ (dotted lines).
Energies are in units of the superconducting gap $\Delta_0$, lengths are in units of the wavelength $\lambda$ of the periodically-modulated field.
}
\label{fig:current}
\end{figure}

As demonstrated, 0D quasi-Majorana modes are pinned to the nodes of the local topological gap $\mathcal M(x)=0$ in the presence of spatially-modulated fields.
This theoretically established the possibility to realize a Majorana chain model in nanowires.
Moreover, this system can be employed to realize an adiabatic ``Majorana pump''.
On top of the field $\mathbf{b}_\text{nm}(x)$ induced by the nanomagnets, let us apply a rotating field in the $zx$-plane forming an angle $\theta$ with the $x$-axis, and a uniform field $-{b}_\text{nm}\hat{\mathbf{z}}$ equal and opposite to the average field of the nanomagnets.
The total field is
\begin{align}\label{eq:rotatingfield}
\mathbf{b}(x)
=
\mathbf{b}_\text{nm}(x)-{b}_\text{nm}\hat{\mathbf{z}}+
b_\text{a}[\cos{\theta}\,\hat{\mathbf{x}} + \sin{\theta}\,\hat{\mathbf{z}}].
\end{align}
If $\mathbf{b}_\text{nm}(x)$ is approximately harmonic as in \cref{eq:field} one has
$|\mathbf{b}(x)|^2=b_\text{nm}^2+b_\text{a}^2-2b_\text{nm}b_\text{a}\sin{(2\pi x/\lambda+\theta)}$,
which gives
$|b|\approx\sqrt{b_\text{nm}^2+b_\text{a}^2}-(b_\text{nm}b_\text{a}/\sqrt{b_\text{nm}^2+b_\text{a}^2})\sin{(2\pi x/\lambda+\theta)}$.
Thus, assuming $\sqrt{\mu^2+\Delta^2}\approx \sqrt{b_\text{nm}^2+b_\text{a}^2}$ the local Majorana mass gap becomes $\mathcal M(x)=(b_\text{nm}b_\text{a}/\sqrt{b_\text{nm}^2+b_\text{a}^2})\sin{(2\pi x/\lambda+\theta)}$ which has equally-spaced nodes at $x_n/\lambda=\theta/2\pi+n/2$.
Slowly varying the applied field direction $\theta$ induces the adiabatic sliding of the 1D lattice of 0D quasi-Majorana modes, corresponding to the pumping of one 0D quasi-Majorana mode every half-turn $\theta\to\theta+\pi$ and one full fermionic state every full turn $\theta\to\theta+2\pi$ of the applied field direction.
In the case of periodic and closed boundary conditions, a half-turn of the applied field direction corresponds to the translation $T$ of 0D quasi-Majorana modes entering the definition of the supercharges in \cref{eq:supercharges}.
\Cref{fig:current}(d) shows the intensity of the total magnetic field in \cref{eq:rotatingfield} as a function of $\theta$ (see also Supplementary Figure 2).
\Cref{fig:current}(a) shows the evolution of the LDOS when the applied field direction $\theta$ turns around in the $zx$-plane.
As the field rotates, 0D quasi-Majorana modes slide along the wire, resulting in an adiabatic pumping of 0D quasi-Majorana modes, as shown in \cref{fig:current}(a).
0D quasi-Majorana modes translate by $T$ at each half-turn $\theta\to\theta+\pi$.
\Cref{fig:current}(b) shows the energy of the dispersive 1D Majorana modes below the gap, which corresponds to the sliding of the lattice of 0D quasi-Majorana modes.
For reference, \cref{fig:current}(c) shows the local Majorana mass $\mathcal M$ and its nodes $\mathcal M=0$ as a function of $\theta$.
This result may be generalized to the dynamical Floquet regime, with the possible realization of a finite ``Majorana current'' through the wire.

\section*{Discussion}

We theoretically proposed the realization of extended quantum-mechanical SUSY with central charges and dispersive 1D Majorana fermions in condensed matter, specifically, in a proximitized semiconducting nanowire via spatially-modulated magnetic fields.
As shown in previous studies, a chain of Majorana modes exhibits both quantum mechanical SUSY~\cite{hsieh_all_2016,huang_supersymmetry_2017} and space-time SUSY~\cite{sannomiya_supersymmetry_2019,rahmani_emergent_2015,rahmani_phase_2015}.
In our work, we unveiled the presence of an additional highly-nontrivial structure, i.e., an extended quantum-mechanical SUSY with central charges.
This structure emerges as the combination of two coexisting $\mathcal N=2$ superalgebras:
To our knowledge, the properties of these two coexisting superalgebras have not been previously demonstrated in the Majorana chain context.
Extended SUSY with central charges has been one of the most important notions in quantum field theory and string theory over the decades since the second string revolution in the 90s and the revolution of quantum field theory by Seiberg and Witten~\cite{Witten:1978mh,Seiberg:1994aj,Seiberg:1994rs}.
In spite of its great importance in formal aspects of quantum field theory and string theory, all high energy theorists regard it as a useful tool which is not directly related to reality.
This is because the extended $\mathcal N=2$ SUSY does not allow chiral fermions relevant for elementary particles such as quarks and leptons.
In high-energy phenomenology, only $\mathcal N=1$ SUSY and its breaking are considered.
There are many proposals to realize SUSY in condensed matter, e.g., in Bose-Fermi mixtures of ultracold atoms~\cite{snoek_ultracold_2005,snoek_theory_2006,yu_supersymmetry_2008,yu_simulating_2010,shi_supersymmetric_2010,lai_relaxation_2015,blaizot_spectral_2015,bradlyn_supersymmetric_2016,blaizot_goldstino_2017,tajima_goldstino_2021}, Majorana Cooper-pair boxes~\cite{ebisu_supersymmetry_2019}, and at the boundaries of topological superconductors or insulators~\cite{grover_emergent_2014,ponte_emergence_2014,ma_realization_2021}.
However, our model describes the first accessible example of extended SUSY with central charges and partial spontaneous SUSY breaking realized in nature.
%We thus believe that they will be shocked to know that the centrally extended SUSY can be realized in real-world physics

The action of the Goldstino is usually accompanied by higher derivative corrections determined thoroughly by a symmetry-breaking pattern, as found by Volkov and Akulov~\cite{volkov_is_1973} (see also recent works on extended SUSY~\cite{cribiori_2d-volkov-akulov_2019,chakrabarti_chiral_2020}).
Our theory should be considered as the leading order of the derivative expansion.
Thus, higher derivative correction terms, if one could obtain them, should be summed up to the Volkov-Akulov type action for 1D Majorana fermions.

While the ${\mathcal N}=4$ SUSY conformal field theories in 1D are known to be characterized by a central charge $c=2$~\cite{fokkema_spinon_2016}, the central charge in our case is $c=1$, due to the presence of two helical pair of counterpropagating Majorana fermions, implying only ${\mathcal N}=2$ SUSY instead of ${\mathcal N}=4$ SUSY\@.
This is compatible with the fact that ${\mathcal N}=4$ SUSY is spontaneously broken and only ${\mathcal N}=2$ SUSY remains, giving strong evidence of the existence of the unbroken extended ${\mathcal N}=2$ SUSY in our model.

We notice that electronic interactions with the external environment may contribute to the pinning of the Majorana modes to zero energy in a typical nanowire setup.
%Generally, spatially-separated Majorana modes localized at the ends of a nanowire with infinite length are charge neutral, since they are an equal superposition of particles and holes.
Without interactions, the SUSY regime corresponds to a single point of the parameter space, which coincides with the closing of the particle-hole gap, and occurs when the externally applied magnetic field is exactly $b_\text{a}= b_\text{SUSY}$.
In the presence of electronic interactions, however, the lowest energy level may become pinned to zero energy, and therefore the particle-hole gap may remain closed in an extended window of the parameter space.
This may occur due to the self-interaction of the charge distribution of the Majorana modes mediated by the external environment~\cite{dominguez_zeroenergy_2017} in the case of a pair of Majorana modes localized at the edges of the nanowire.
Perfectly spatially-separated Majorana modes are charge neutral.
However, in our case, contiguous 0D quasi-Majorana modes have a finite overlap, which gives rise to a finite charge density~\cite{lin_zerobias_2012,ben-shach_detecting_2015}.
This finite charge induces a screening charge distribution in the dielectric environment, which acts back onto the Majorana modes.
This results in a self-interaction term that pushes the energy of the Majorana modes back to zero, as long as the nanowire has a larger dielectric constant of the external environment (e.g., the nanowire substrate)~\cite{dominguez_zeroenergy_2017}.
This mechanism can stabilize the SUSY by pinning the Majorana mass $\mathcal M_\text{eff}\equiv0$ over an extended parameter space.

To experimentally realize our proposal, the wire must be much longer than the field periodicity, which must be comparable with the Majorana localization length, i.e., $L\gg\lambda\gtrsim\xi_\text{M}{\approx\alpha/\Delta}$.
Moreover, variations of the gate and spin-orbit coupling fields must be negligible at length scales larger than $\lambda$, to guarantee an unbroken translational invariance at the mesoscopic level.
Conversely, the physics described here is not affected by perturbations having a length scale shorter than $\lambda$ (e.g., disorder).
A different approach to realizing localized quasi-Majorana modes is by employing noncollinear magnetic textures or domain walls in complex magnet-superconductor heterostructures~\cite{neupert_chain_2010,rex_majorana_2020}.
However, our proposal does not require the presence of a magnetic substrate and has the advantage of using proximitized semiconducting nanowires, which by far are the most extensively studied platform for topological superconductivity~\cite{stanescu_majorana_2013,lutchyn_majorana_2018,zhang_next_2019,frolov_topological_2020,flensberg_engineered_2021}.

In a finite wire with $2N$ 0D Majorana modes with open boundary conditions and at zero temperature, the differential conductance exhibits $2N-1$ zeros and $2N$ quantized peaks $G=2e^2/h$, and the zero-bias conductance is zero~\cite{flensberg_tunneling_2010}.
In an infinite wire with identical couplings ($v=w$) instead, the conductance shows a zero-bias peak $G=2e^2/h$~\cite{flensberg_tunneling_2010}, which corresponds to the tunneling into the delocalized Majorana mode at zero energy.
For $v\neq w$, the energy of the Majorana modes is lifted by the broken SUSY, and therefore there is no Majorana mode available at zero energy: In this case, the zero-bias conductance is zero.
Hence, for a sufficiently long Majorana chain or equivalently in the case of closed boundary conditions (i.e., in a loop geometry), the transition between a zero-bias dip $G=0$ to a peak $G=2e^2/h$ signals the onset of SUSY at $v=w$.
This transition should be observable in sufficiently long nanowires or in a setup with loop geometry (i.e., with closed boundary conditions) by varying the applied magnetic field close to the SUSY point $b_\text{a}= b_\text{SUSY}$.
However, these signatures may be difficult to distinguish from conductance peaks induced by disorder, impurities, or finite-size effects.
The characterization of the signatures of SUSY in the conductance will be the subject of future work.
Stronger experimental signatures are the presence of spatially-periodic peaks in the LDOS along the whole wire, and their adiabatic evolution in the Majorana pump regime, obtained by varying the applied magnetic field direction.
These signatures can be obtained by locally probing the differential conductivity in multiterminal setups~\cite{grivnin_concomitant_2019,menard_conductance-matrix_2020,puglia_closing_2021,heedt_shadow-wall_2021}, or by scanning tunneling microscopy (STM) in epitaxial 1D heterostructures~\cite{shabani_two-dimensional_2016,hell_two-dimensional_2017,pientka_topological_2017,suominen_zero-energy_2017}.
%Another possible signature can be provided by the adiabatic Majorana pump regime, realized by varying the applied magnetic field direction, by a local probe of the variations of the LDOS induced by the sliding lattice of 0D quasi-Majorana modes.
Other nonlocal fingerprints of SUSY can be provided by the signatures of the closing of the bulk particle-hole gap with a finite density of states on the whole wire.
In particular, these signatures can be revealed via a tunneling probe placed at the bulk of the nanowire~\cite{grivnin_concomitant_2019}, by the quantized peak of the thermal conductance and electrical shot noise at the transition, and the doubling of the magnetoconductance oscillations in an Aharonov-Bohm ring geometry~\cite{akhmerov_quantized_2011}, or by the peak of the $4\pi$ component of the Josephson current in a superconducting ring geometry~\cite{pientka_signatures_2013}.

We also mention that the proposed experimental protocol to realize an adiabatic Majorana pumping in a sliding lattice of 0D quasi-Majorana modes may suggest new methods of braiding Majorana modes in 1D nanowire networks, which is the next milestone in the route to topological quantum computation~\cite{ivanov_nonabelian_2001,kitaev_faulttolerant_2003,nayak_nonabelian_2008,alicea_nonabelian_2011,sarma_majorana_2015,lahtinen_short_2017,lian_topological_2018}.

Concluding, we proposed an experimentally accessible realization of a Majorana chain with emergent SUSY and with dispersive 1D Majorana fermions in proximitixed nanowires via spatially-modulated magnetic fields.
In this system, we demonstrated the presence of an extended $\mathcal N=4$ SUSY with central charges and we identified the massless 1D Majorana fermions as the Nambu-Goldstone fermions (Goldstinos) associated with the spontaneous partial breaking of SUSY\@.
Their experimental signatures are the finite LDOS at zero energy (zero-bias peak) delocalized on the whole length of the wire, and the dip-to-peak transition in the zero-bias conductance.
This has to be contrasted with zero-bias peaks of 0D Majorana end modes, localized only at the ends of the wire, and to the general case of 0D quasi-Majorana modes, whose energy is lifted by their finite overlap.
We finally showed how to realize an adiabatic Majorana pump by varying the applied magnetic field direction, which induces a sliding lattice of 0D quasi-Majorana modes with quantized transport of one quasi-Majorana mode per a half cycle.
The manipulation of Majorana modes via spatially-modulated fields may lead to the realization of alternative non-abelian braiding protocols.

\section*{Methods}
The numerical results were obtained by discretizing \cref{eq:LEH} into a tight-binding model and diagonalizing the resulting Hamiltonian.
The LDOS in the main article and in the Supplementary Information was calculated from the spectra as a function of energy.
The momentum dispersion was obtained by Fourier-transform the tight-binding Hamiltonian and diagonalizing the resulting Hamiltonian in the momentum basis.
In agreement with previous works~\cite{mourik_signatures_2012,liu_andreev_2017}, we consider an InSb nanowire proximitized by a conventional superconductor, and take $m=0.015\,m_\text{e}$, $\alpha=\text{\SI{1}{\eV\angstrom}}$, $b/B=\text{\SI{1.5}{\milli\eV\per\tesla}}$ ($g\approx50$), $B_\text{c}=\text{\SI{3}{\tesla}}$, and $\Delta_0=\text{\SI{1}{\milli\eV}}$ at zero magnetic field.
In this regime, the Majorana localization length is estimated to be $\xi_\text{M}\approx\text{\SIrange{200}{300}{\nano\metre}}$ depending on the applied field $b_\text{a}$.
The magnetic field induced by the array of nanomagnets was calculated by numerical integration over a finite mesh.
We considered a periodic array of \SI{250}{\nano\metre}$\times$\SI{250}{\nano\metre}$\times$\SI{250}{\nano\metre} nanomagnets placed at a mutual distance \SI{750}{\nano\metre} and at $d=\text{\SI{250}{\nano\metre}}$ from the wire, with remnant magnetic field $B_\text{R}\approx\text{\SI{1}{\tesla}}$ and parallel magnetic moments in the $z$ direction.
See Supplementary Note 1 and Supplementary Figures 1 and 2 for more details.

\section*{Data availability}
The code used for the numerical simulations within this paper and the resulting data are available on Zenodo~\cite{marra_data_2022}.

%\section*{Acknowledgements}
\begin{acknowledgments}
P.~M.~thanks Sven Bjarke Gudnason, Stefan Rex, Masatoshi Sato, and Benjamin Woods for useful suggestions.
P.~M.~is supported by the Japan Science and Technology Agency (JST) of the Ministry of Education, Culture, Sports, Science and Technology (MEXT), JST CREST Grant.~No.~JPMJCR19T2, the MEXT-Supported Program for the Strategic Research Foundation at Private Universities ``Topological Science'' (Grant No.~S1511006), and Japan Society for the Promotion of Science (JSPS) Grant-in-Aid for Early-Career Scientists (Grant No.~20K14375).
D.~I.~is supported by the Financial Support of Fujukai Foundation.
M.~N.~is partially supported by the JSPS Grant-in-Aid for Scientific Research (KAKENHI) Grant Number 18H01217.
\end{acknowledgments}

\section*{Author contributions}
P.~M. carried out the numerical calculations.
P.~M., D.~I., and M.~N. contributed to the scientific discussion and writing of the manuscript.

\section*{Competing interests}
The authors declare no competing interests.

%\bibliography{bib,other}
%\bibliographystyle{prsty_no_etal_titles_doi_preprint_noemph}

\end{document}